# Prospects of Direct Growth Boron Nitride Films as Substrates for Graphene Electronics

Michael S. Bresnehan,[1,2,3] Matthew J. Hollander,[3,4] Maxwell Wetherington,[1,2,3] Ke Wang,[5] Takahira Miyagi,[1,5] Gregory Pastir,[2] David W. Snyder,[2,6] Jamie J. Gengler,[7,8] Andrey A. Voevodin,[7] William C. Mitchel,[7] and Joshua A. Robinson[1,3,*]

[1]Department of Materials Science and Engineering, [2]Electro-Optics Center, [3]Center for 2-Dimensional and Layered Materials, [4]Department of Electrical Engineering, [5]Materials Research Institute, and [6]Department of Chemical Engineering; The Pennsylvania State University, University Park, Pennsylvania 16802, United States. [7]Materials and Manufacturing Directorate, Air Force Research Laboratory, WPAFB, OH 45433. [8]Spectral Energies, LLC, Dayton, Ohio 45431, United States.

*jrobinson@psu.edu

**ABSTRACT**

We present a route for direct growth of boron nitride *via* a polyborazylene to h-BN conversion process. This two-step growth process ultimately leads to a >25x reduction in the RMS surface roughness of h-BN films when compared to a high temperature growth on $Al_2O_3$(0001) and Si(111) substrates. Additionally, the stoichiometry is shown to be highly dependent on the initial polyborazylene deposition temperature. Importantly, CVD graphene transferred to direct-grown boron nitride films on $Al_2O_3$ at 400°C results in a >1.5x and >2.5x improvement in mobility compared to CVD graphene transferred to $Al_2O_3$ and $SiO_2$ substrates, respectively, which is attributed to the combined reduction of remote charged impurity scattering and surface roughness scattering. Simulation of mobility versus carrier concentration confirms the importance of limiting the introduction of charged impurities in the h-BN film and highlights the importance of these results in producing optimized h-BN substrates for high performance graphene and TMD devices.

## Table of Contents / Abstract Image

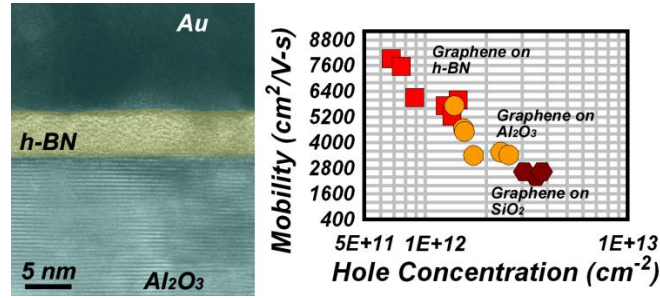

**Introduction**

Hexagonal boron nitride (h-BN) has attracted considerable attention as a complementary dielectric material in graphene based electronics.[1–5] Owing to its sp[2] hybridized bonding and weak inter-planar van der Waals bonds, h-BN is expected to be free of dangling bonds.[1,6] This is expected to lead to a decreased density of absorbed impurities that can act as Coulombic scatterers in graphene-based electronics.[6] Additionally, compared to high-k dielectrics, h-BN has high energy surface optical phonon modes that allow for reduced scattering from electron-phonon interactions at the graphene/dielectric interface.[1] Indeed, exfoliated h-BN and transferred CVD h-BN have been shown to be excellent candidates as a supporting substrate and/or gate dielectric to CVD graphene,[2–5] as well as a top-gate dielectric to epitaxial graphene.[7,8] Besides graphene, h-BN is of interest as a substrate for development of other 2D layered materials, such as transition metal dichalcogenides (TMDs),[9] where h-BN's close lattice match makes it an ideal candidate as a substrate for growth and integration. Despite the advantages of h-BN over high-k dielectrics and $SiO_2$, exfoliated h-BN is not suitable for large scale development, and CVD of h-BN on metal substrates requires a transfer process that can be difficult to perform with repeatable success.[10] This transfer process often results in a rough h-BN surface due to stress induced wrinkling of the film during growth on the transition metal surface, where surface roughness and wrinkling increase with film thickness (see Figures S1 and S2 of the supplemental information).[7] Additionally, the formation of tears and pinholes can easily occur during the transfer process, leading to difficulty in producing continuous films over large areas. Contamination issues can also arise during the solution-based transfer process, usually in the form of residual etchant such as iron from the ferric chloride solution used to etch copper and nickel.[7] Therefore, a deposition process that can grow high-quality, ultra-smooth, transfer-free h-BN with controllable thicknesses directly on a variety of insulating and conductive supporting substrates would be highly beneficial for graphene and TMD development. Here, we present a direct deposition process of h-BN on $Al_2O_3$(0001) and conductive Si(111) that involves the initial deposition of the hydrogenated polymer precursor polyborazylene ($B_3N_3H_x$) and its subsequent conversion to boron nitride. We discuss the effects of the initial polyborazylene deposition temperature on the chemical composition as measured by X-ray photoelectron spectroscopy (XPS), the surface morphology and roughness as characterized by atomic force microscopy (AFM), and the film structure as examined by cross-sectional transmission electron microscopy (TEM) for h-BN films grown directly on $Al_2O_3$(0001) and Si(111) substrates. Additionally, we discuss thermal properties of the h-BN interfaces and their implications on device reliability. Finally, the effect of the h-BN supporting substrates on the transport properties of transferred CVD graphene was examined *via* Hall mobility measurements. The results show that by optimizing the conditions for direct growth of h-BN, up to a >2.5x increase in mobility can be achieved for CVD graphene transferred to these substrates. These results highlight the importance of tailoring the growth conditions to limit the introduction of charged impurities and surface roughness, which are highlighted as the two main scattering processes leading to degradation in transport properties.

**RESULTS AND DISCUSSION**

Initial attempts to deposit h-BN directly on Si, SiO$_2$, and sapphire using high temperature growth conditions (1000°C) resulted in thick h-BN films with a high surface roughness not suitable for device development. Therefore, a two-step deposition process, similar to the 2-step growth of h-BN on Ni substrates reported elsewhere,[11] was developed to reduce surface roughness. The process consists of depositing a film of polyborazylene (from sublimation of ammonia borane) onto a substrate at low temperatures followed by a high temperature anneal at 1000°C to convert the polyborazylene to h-BN.[11] The deposition of polyborazylene was performed under 100 Torr of H$_2$/N$_2$ (15% H$_2$) carrier gas at two temperatures, 250°C and 400°C. These two temperatures were selected to represent a temperature within the first hydrogen loss regime of polyborazylene and one slightly above the regime. A post-growth anneal was then performed at 1 Torr H$_2$/N$_2$ (15% H$_2$), 1000°C for one hour to facilitate the conversion to boron nitride.

The surface morphology of direct-growth h-BN depends critically on the deposition and anneal conditions of the polyborazylene. High temperature growth (1000°C, 100 Torr, 10 minutes) of h-BN on Si and Al$_2$O$_3$ without the use of an initial polyborazylene deposition (Figure 1a and b) results in a 25 nm thick film with a root-mean-square (RMS) roughness >3 nm. Utilizing an initial polyborazylene deposition at 400°C (400°C, 100 Torr, 10 minutes) followed by a post-growth anneal at 1000°C (Figure 1c and d) yields a reduced film thickness of 8 nm and a significant reduction in surface roughness of >13x (RMS roughness = 0.24 nm) compared to the single-step high temperature growth process. Additionally, a >16x decrease in RMS roughness (4.02 nm to 0.24 nm) was found when this sample was compared to an h-BN film of similar thickness grown via CVD on a copper substrate and transferred to Si(111) (see supplemental information, Figure S2d). Further reducing the polyborazylene deposition temperature to 250°C (250°C, 100 Torr, 120 minutes) yields a 2 nm thick film with an RMS roughness of 0.13 nm on the

Al$_2$O$_3$ sample. Here, a >25x reduction in RMS roughness (3.28 nm to 0.13 nm) was observed compared to the high temperature deposition (Figure 1e) and a >9x reduction (1.18 nm to 0.13 nm) compared to transferred h-BN of similar thickness (see supplemental information, Figure S2b). Interestingly however, when the same conditions are used for h-BN growth on Si(111), partial film voiding occurs (Figure 1f). This has been observed previously for spin coated polyborazylene on Si after high temperature annealing and was attributed to degradation of the underlying native oxide.[12] Therefore, a pre-growth anneal at 1000°C and 100 Torr in an H$_2$/N$_2$ environment was performed to remove any native oxide on the Si surface prior to growth at 250°C.[13] This resulted in an increased RMS roughness of the film to approximately 1 nm (Figure 1g and h). This roughening may result from several possible mechanisms. One possibility is that an initial deposition of BN during the 1000°C pre-growth anneal may occur due to desorption and re-deposition of residual BN from the furnace walls; acting as a seed layer with a rough morphology, as similarly observed for the 1000°C BN deposition shown in Figures 1a and b. An alternative possibility is that the Si and Al$_2$O$_3$ substrate surfaces may reconstruct and roughen during the 1000°C pre-growth anneal. This roughening of the substrate surface may translate to the morphology of the deposited polyborazylene film. Although further analysis is required to fully understand the surface roughening resulting from the high temperature pre-growth anneal, what can be concluded is that the film voiding on the Si sample was successfully eliminated through use of this high temperature pre-growth treatment (Figure 1h).

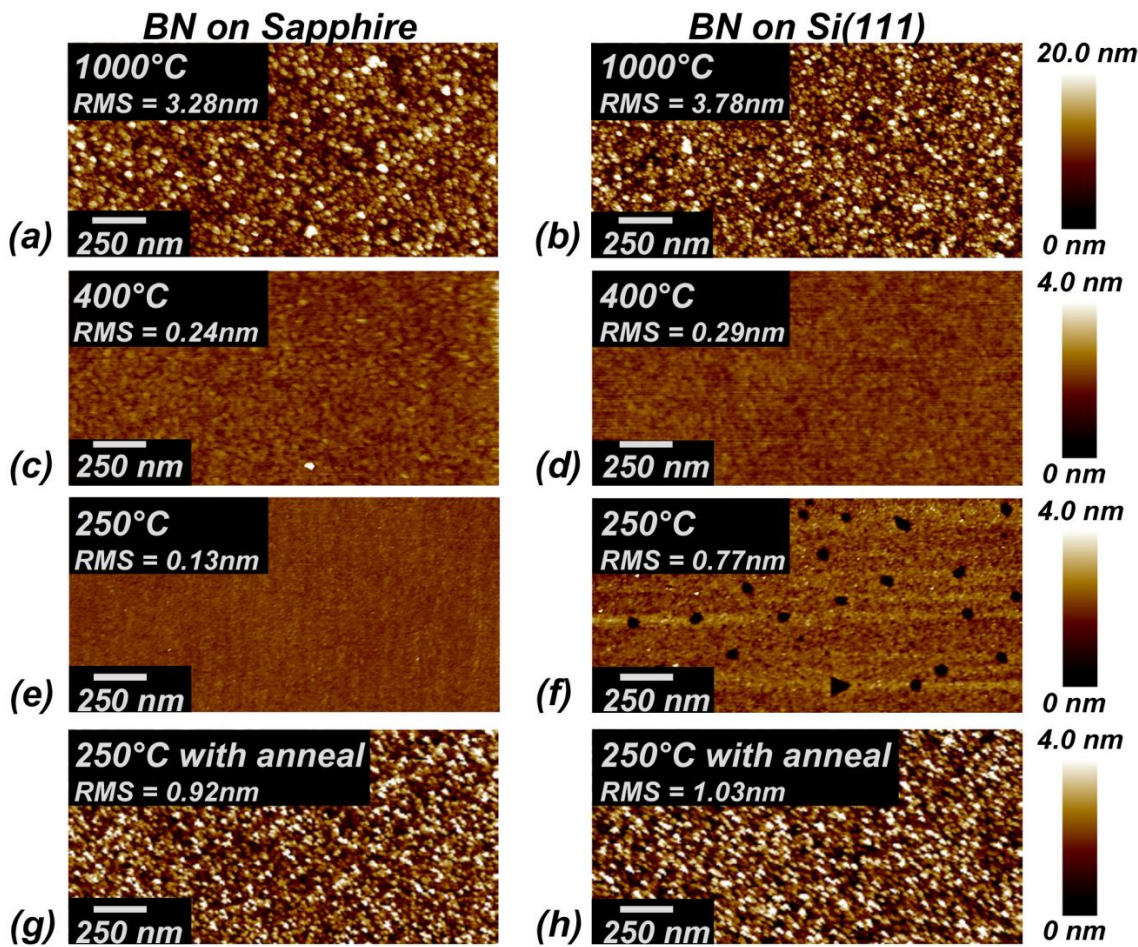

**FIG. 1:** AFM scans of h-BN directly grown on $Al_2O_3$(0001) and Si(111) at various deposition conditions. (a,b) Growth of h-BN on $Al_2O_3$ and Si, respectively, at 1000°C without the use of an initial polyborazylene deposition. (c,d) Growth of h-BN on $Al_2O_3$ and Si, respectively, using an initial polyborazylene deposition temperature of 400°C. (e,f) Growth of h-BN on $Al_2O_3$ and Si, respectively, using a polyborazylene deposition temperature of 250°C. (g,h) Growth of h-BN on $Al_2O_3$ and Si, respectively, using an initial polyborazylene deposition temperature of 250°C and a pre-growth anneal of 1000°C. Note that the scale bar for (a, b) is 20 nm while the scale bar for (c-h) is 4 nm.

In addition to surface morhology of direct grown h-BN, the chemical state of the films is also critical to understand because this can lead to charge impurity scattering in graphene overlayers, and ultimately reduce performance of graphene-based devices. X-ray photoelectron spectroscopy (XPS) was used to investigate the bonding of boron nitride grown at different polyborazylene deposition temperatures. Figure 2 shows the deconvoluted B 1s spectra of h-BN grown on $Al_2O_3$ substrates using the four growth conditions described previously. The binding energies of all spectra were obtained after charge correction of the C 1s line to adventitious carbon at 284.8 eV. Except where noted, the chemistries of the h-BN samples grown on Si(111) are similar to the h-BN samples on $Al_2O_3$(0001). Initial studies focused on direct growth at 1000°C (Figure 2a), which yields a mixture of B-N-O bonding. The primary peak at 190.25 eV, accounting for 69.8% of the main B 1s line, is indicative of the h-BN structure where the core boron atom is bonded to three nitrogen atoms in the planar hexagonal configuration and can be written as [$BN_3$].[14] The smaller shoulder at 191.01 eV indicates the presence of oxygen and occupies 30.2% of the main B 1s line. Pure $B_2O_3$ has been previously reported to be located between 193.0-193.6 eV.[15–17] This intermediate peak between the h-BN [$BN_3$] peak and the reported $B_2O_3$ peak was similarly observed by Guimon *et al.*[15] and was attributed to the core boron atom bonding to nitrogen and oxygen, where oxygen may be a substitutional element in place of nitrogen, forming a $BN_xO_y$ ternary species. Since an ultra-high vacuum system was not used for this study, oxygen incorporation is likely due to oxygen impurities in the gas phase and has been commonly reported in the literature.[15,17] Additionally, the π-π* plasmon shake-up peak at 198.62 eV is also present, corresponding to $sp^2$ bonded hexagonal boron nitride.[17,18]

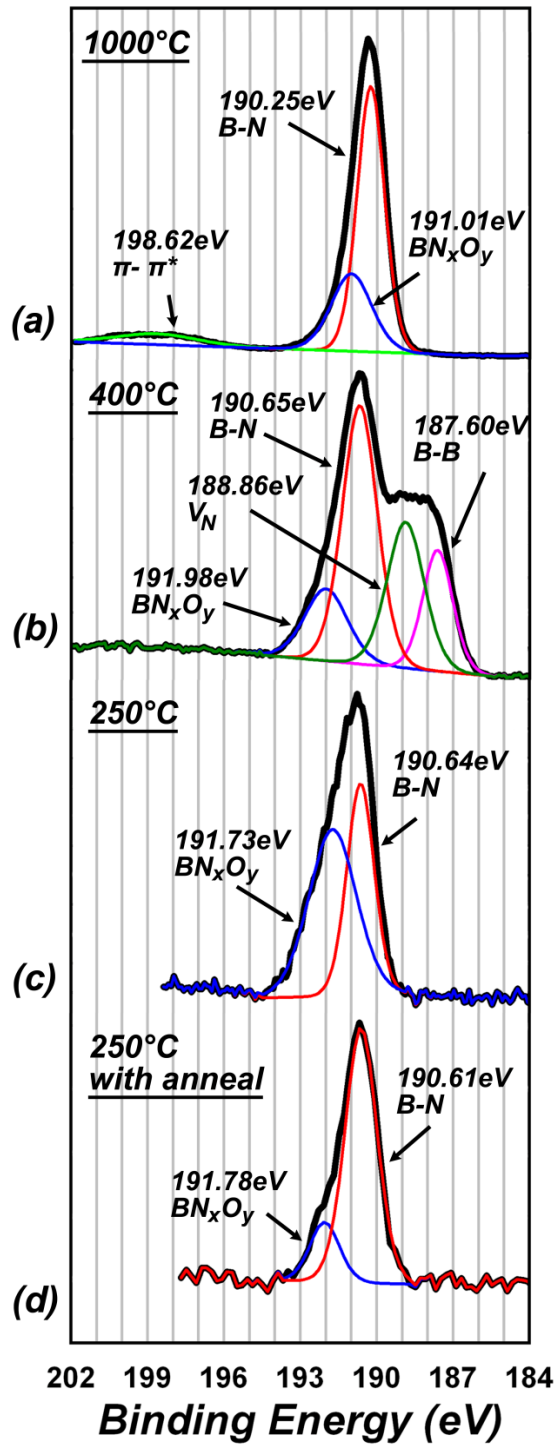

**FIG. 2:** XPS B 1s spectra of h-BN directly grown on $Al_2O_3(0001)$ at (a) 1000°C, (b) 400°C, (c) 250°C, and (d) 250°C with a 1000°C pre-growth anneal. Component peaks were deconvoluted using a Gaussian fit. Peak positions are given after performing a charge correction of the C 1s core level peak to 284.8 eV.

Polyborazylene deposition at 400°C, while yielding a smooth surface morphology, leads to loss of stoichiometry (Figure 2b). XPS reveals that there are four distinct chemical bonding phenomenon, as evident after deconvolution of the B 1s peak. The primary [BN$_3$] bonding configuration is found at 190.65 eV and accounts for 42.7% of the B 1s spectrum. Additionally, two shoulders at energies lower than the [BN$_3$] peak are found at 188.86 eV and 187.60 eV. Elemental boron-boron bonding has been reported previously in the range of 187.3-187.7 eV.[14,19] Here, the peak at 187.60 eV accounts for 17.2% of the B 1s spectrum and can be attributed to elemental boron with no bonds to nitrogen atoms due to boron-boron bonding in either interstitial or interlayer positions, or at defective domain boundaries. Bonding with nitrogen atoms creates a higher binding energy compared to elemental boron-boron bonding due to the core boron atom bonding with the higher electronegative element nitrogen.[14] Therefore, the peak inbetween the h-BN [BN$_3$] and the B-B configuration, located at 188.86 eV (25.7% of the B 1s spectrum), can be attributed to either a [BN$_2$] configuration, with boron bonded to two nitrogen atoms, or a [BN] configuration with boron bonded to one nitrogen atom. In other words, the intermediate peak corresponds to a single or double nitrogen vacancy ($V_N$). A similar peak was reported by Guimon *et al.*, where after 4 keV Ar$^+$ sputtering, a peak at 189.2 eV appeared.[15] Likewise, Schild *et al.* reported the emergence of a peak at 188.8 eV after sputter cleaning of c-BN samples.[17] In both cases this peak was attributed to preferential sputtering of nitrogen atoms, leaving behind a boron-rich surface. Additionally, a peak at 191.98 eV again is attributed to substitutional oxygen contamination in the form of a BN$_x$O$_y$ ternary species and accounts for 14.5% of the B 1s spectrum. Due to the elemental boron bonding, nitrogen vacancies, and substitutional oxygen bonding, a B/N ratio of 1.92 was obtained for this sample, indicating a highly boron-rich film. Interestingly, the presence of B-B bonding in this film may result in a reduction of free electrons in the material by compensating for the difference in valence between nitrogen and oxygen, where the oxygen impurities will induce additional valence electrons.

Therefore, despite the fact that this film is highly non-stoichiometric, the B-B bonding may act to create a more charge-neutral dielectric material (compared to direct-growth BN films containing oxygen contamination but no B-B bonding) which may result in reduced electrostatic doping to graphene.

Polyborazylene deposition at 250°C without the use of a pre-growth anneal results in the presence of two types of bonding, assigned to h-BN [$BN_3$] (190.64 eV) and $BN_xO_y$ (191.73 eV). Interestingly, for this sample the $BN_xO_y$ peak accounts for 57.7% of the B 1s spectra, indicating significant oxygen incorporation. Electron energy loss spectroscopy (EELS) (not shown) indicates that for the majority of these samples, interdiffusion between the substrate and the h-BN layer occurs up to 3 nm. Given that this sample is approximately 2 nm, the additional oxygen incorporation is likely induced from the $Al_2O_3$ substrate and substitutional bonding occurs during the high temperature anneal. The nitrogen vacancy peaks and B-B bonding peak (observed in the film grown at 400°C) are not observed for this sample, possibly due to oxygen occupying vacancy sites. However, the additional substitutional oxygen contamination results in a B/N ratio of 1.51. Similarly, a polyborazylene deposition temperature of 250°C *with* the use of a 1000°C pre-growth anneal (Figure 2d) resulted in two types of bonding, assigned to h-BN [$BN_3$] (190.61 eV) and $BN_xO_y$ (191.78 eV). However, here the $BN_xO_y$ peak is significantly reduced compared to the sample grown at 250°C *without* a pre-growth anneal; accounting for 17.7% (compared to 57.7%) of the B 1s spectrum, despite the fact that these films were of similar thickness (~2 nm). This may be attributed to the 1000°C pre-growth anneal, where hydrogen from the $H_2/N_2$ (15% $H_2$) environment may act to reduce the $Al_2O_3$ surface, leaving an oxygen-deficient interface for polyborazylene deposition. This would intuitively result in a reduced interdiffusion of oxygen into the BN film. Alternatively, a BN seed layer may be deposited during the 1000°C pre-growth anneal that acts to reduce interdiffusion between the $Al_2O_3$ substrate and the h-BN layer. Unlike the sample grown at 400°C however, the lower energy peaks corresponding to nitrogen vacancies and B-B bonding are absent and the B/N ratio was found to be 1.01, indicating excellent stoichimetry. This may again be a result of a BN seed layer grown during the high temperature anneal, which may act to promote further BN bonding and reduce B-B bonding. Further analysis is required, however, to fully understand the mechanisms responsible for interdiffusion between BN and substrate.

Table I shows the B, N, O, and C atomic percentages and B/N ratios for h-BN grown on both $Al_2O_3$(0001) and Si(111) for all four growth conditions. In three of the four cases, the atomic concentrations and B/N ratios are comparable between both substrates. However, the sample grown at 250°C without the use of a pre-growth anneal shows significant variance between the films grown on $Al_2O_3$ and Si. Unlike the sample grown on $Al_2O_3$, the film grown on Si shows significantly reduced oxygen contamination and excellent stoichiometry with a B/N ratio of 1.00. Clearly, oxygen contamination from the substrate is reduced with use of an Si substrate; additionally, the overall oxygen content is reduced in comparison to the other h-BN films grown on Si. Due to the small thickness of this sample (~2 nm), it is possible that oxygen from the gas-phase environment diffuses through the BN layer and reacts with the Si substrate to form $SiO_2$, rather than reacting with the newly formed polyborazylene film. The formation of this $SiO_2$ layer may also explain the pitting observed with AFM in Figure 1f; where during the 1000°C post-growth anneal, this $SiO_2$ layer decomposes and results in surface pitting.

**TABLE I.** Atomic concentrations of boron, nitrogen, oxygen, and carbon and B/N ratios as measured with XPS of h-BN films grown at various conditions on $Al_2O_3(0001)$ and Si(111).

| h-BN on $Al_2O_3$(0001) | B | N | O | C | B/N |
|---|---|---|---|---|---|
| 1000°C | 46.9% | 43.8% | 6.2% | 3.2% | 1.1 |
| 400°C | 59.5% | 31.1% | 5.9% | 3.5% | 1.9 |
| 250°C | 24.3% | 15.7% | 50.1% | 9.9% | 1.6 |
| 250°C with anneal | 38.5% | 38.0% | 19.6% | 3.9% | 1.0 |
| h-BN on Si(111) | B | N | O | C | B/N |
| 1000°C | 46.1% | 41.5% | 7.1% | 5.4% | 1.1 |
| 400°C | 56.0% | 33.0% | 7.1% | 3.9% | 1.7 |
| 250°C | 42.6% | 42.5% | 4.8% | 10.1% | 1.0 |
| 250°C with anneal | 41.6% | 37.8% | 16.3% | 4.4% | 1.1 |

The exact mechanisms for the variance in stoichiometry and bonding configurations between different growth conditions is not fully understood, although boron-rich stoichiometries have been reported previously for boron nitride films synthesized from polyborazylene.[12,20] To rule out the effect of variance in gas-phase chemistry on film stoichiometry, mass spectrometry was used. Here, the concentrations of various gas-phase BN precursors (produced from sublimation of ammonia borane at 135°C) were obtained for furnace temperatures of 250, 400, and 1000°C. The results, shown in the supplemental material (Figure S5), show that the concentrations of borazine, diborane, aminoborane, and $BH_2$ remain independent of temperature. Therefore, the observed differences in stoichiometry at these growth temperatures is not due to the gas-phase chemistry in the CVD environment. Instead, nitrogen may be lost during the high temperature annealing process required to form BN. Indeed, Chan *et al.* demonstrated an increase in the B/N ratio, from 1.18 to 1.37, from spin coated polyborazylene to BN upon annealing to 900°C.[12] It is broadly accepted that the formation of BN from polyborazylene at high temperatures results in the loss of hydrogen during cross-linking of unaligned chain branch structures.[21] It is possible that some B-B bonding may occur during this process, where the bonding energy of B-B bonds has been reported to be 38% less than that of B-N bonds (310 and 500 KJ/mol, respectively),[22] resulting in a loss of nitrogen locally and the formation of a boron-rich film.

The nanoscale structure of h-BN is also influenced by growth parameters. Cross-sectional TEM was utilized to characterize the cross-sections of h-BN films grown on $Al_2O_3$ substrates (structures on Si(111) are similar). For comparison, a ~5 nm h-BN film grown on Cu foil and transferred to $Al_2O_3$ is shown in the supplemental material (Figure S6). The sample grown directly at 1000°C is shown in Figure 3a and suggests that the initial h-BN layers grow nearly parallel to the $Al_2O_3$ surface. After ~5 nm however, the c-axis orientation of the h-BN layers becomes highly random, and is likely the cause of the rough film surface. The film thickness is ~25 nm after only 10 minutes of growth, indicating a comparitively large growth rate of 2.5 nm/min. Additionally, the film is polycrystalline with domain sizes <5 nm. The small domain size is likely a result of stress induced in the h-BN layers due to both the mismatch in coefficient of thermal expansion (CTE) between BN and the substrate (where h-BN has a negative CTE, leading to expansion and compressive stress incorporation in the BN film upon cooling),[7] and to the large lattice mismatch between $Al_2O_3$ and h-BN of 46%, where $Al_2O_3$ and h-BN have c-plane lattice constants of 4.67Å and 2.50 Å, respectively.[23] When polyborazylene is deposited at 400°C (Figure 3b), a ~5 nm film was obtained after 10 minutes of growth indicating a growth rate of 0.5 nm/min. This film appears largely amorphous, although some layering of nano-crystalline domains can be observed, indicating a turbostratic structure. For the films grown at 250°C without a pre-growth anneal (Figure 3c), and 250°C with a pre-growth anneal (Figure 3d), the crystallinity is decreased even further and the films appear amorphous with no long range order. The poor crystallinity obtained in the films grown with the initial polyborazylene deposition can be attributed to the nature of the initial polyborazylene film, which consists primarily of unaligned polyborazylene chain branches that likely bond with only short-range order upon conversion to BN.[24] In arreement with Chan et al.[12] and Kho et al.[20], EELS (not shown) indicates that h-BN films grown at 250°C without a post-growth anneal (Figure 3c) resulted in significant oxygen incorporation arising from interdiffusion from the substrate, also in agreement with XPS (Figure 2c). This interdiffusion layer was found to be present in all four samples, where aluminum and oxygen from the $Al_2O_3$ substrate diffused into the h-BN layers. For the thicker films, grown at higher temperatures (Figures 3a and b), the interface region was found to be ~3 nm in width. This indicates that for the ~2 nm thick samples (Figures 3c and d), significant fractions of the h-BN film are likely to be $BN_xO_y$, corresponding to XPS measurements showing the atomic concentrations of oxygen to be higher in these films (Table I). However, EELS shows somewhat reduced incorporation of oxygen in the sample grown at 250°C with the use of a pre-growth anneal, indicating that a BN seed layer or hydrogen reduction of $Al_2O_3$ during the pre-growth anneal may act to partially reduce oxygen interdiffusion, in agreement with the XPS results described previously (Figure 2d). Similar results were observed for the films grown on Si(111) in this study, where Si was interdiffused in the BN layer.

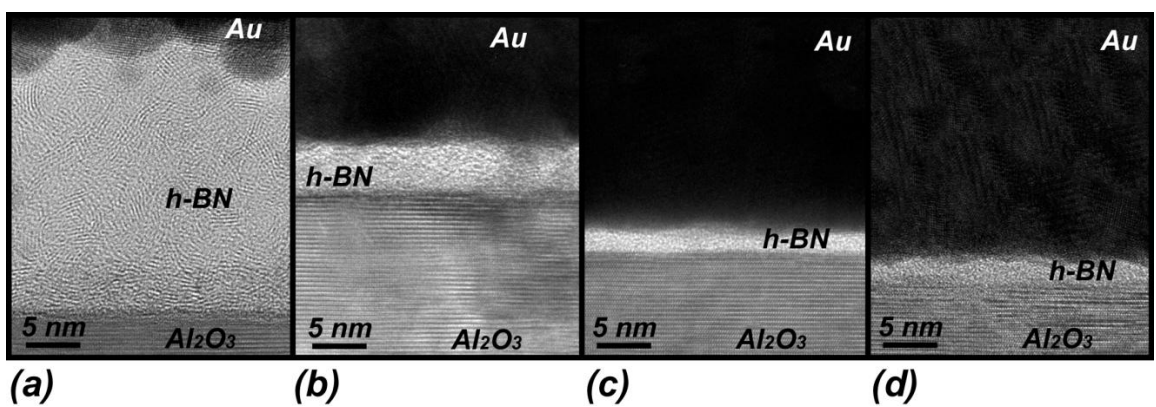

**FIG. 3:** Cross sectional TEM micrographs of h-BN films grown on $Al_2O_3$ at (a) 1000°C without an initial polyborazylene deposition, (b) a polyborazylene deposition temperature of 400°C, (c) a polyborazylene deposition temperature of 250°C without the use of a pre-growth anneal, and (d) a polyborazylene deposition temperature of 250°C and with the use of a 1000°C pre-growth anneal.

To evaluate these films for long term prospects for applications in graphene-based or additional electronic systems, it is instructive to understand their influence on electronic and thermal properies of the heterostructure system.  Therefore, h-BN films were investigated for their effective thermal interface conductance with the time-domain thermoreflectance (TDTR) method,[25] which is sensitive to the inter-atomic bonding and ordering in the thin-film structure as well as the h-BN roughness. Table II summarizes results of the TDTR measurements, while data fitting examples are provided in supplemental Figure S7. For thicker h-BN films produced by a direct one-step growth at 1000°C, it was possible to extract the film cross-plane thermal conductivity. It was interesting to observe for these films that the h-BN thermal conductivity was very low. Conductivities of about 600 $Wm^{-1}K^{-1}$ along base planes and 30 $Wm^{-1}K^{-1}$ in cross base plane directions are normally expected for bulk h-BN crystals. For thin ordered h-BN, these are reduced by a factor of two, where recent evaluations of suspended exfoliated few layer h-BN films had shown 250 $Wm^{-1}K^{-1}$ for in-plane thermal conductivity of 5ML h-BN films.[26] The results of our TEM studies show that h-BN films synthesized from 1000°C direct-growth are nanocrystalline and randomly oriented, which critically influences their overall thermal conductivity. Thermal conductivity values reported by Muratore, C. *et al.*[27] for nanocrystalline hexagonal thin films of $MoS_2$, $WS_2$, and $WSe_2$ indicate that several orders of magnitude reduction in thermal conductivity can be expected in disordered hexagonal structures with weak cross-plane bonding characteristics.

**TABLE II.** Results of TDTR studies of *h*-BN films directly grown on silicon and sapphire substrates at different growth conditions.

| Substrate | Growth Condition | Thickness (nm) | Interface Thermal Conductance (MWm$^{-2}$K$^{-1}$)* | BN Thermal Conductivity (Wm$^{-1}$K$^{-1}$) |
|---|---|---|---|---|
| Si(111) | 1000°C | 25 | 17 ± 1 | 0.46 ± 0.01 |
| Si(111) | 400°C | 7 | 51 ± 2 | - |
| Si(111) | 250°C | 2 | 65 ± 2 | - |
| Si(111) | 250°C with anneal | 2 | 72 ± 2 | - |
| Al$_2$O$_3$(0001) | 1000°C | 25 | 20 ± 1 | 0.57 ± 0.02 |
| Al$_2$O$_3$(0001) | 400°C | 7 | 58 ± 1 | - |
| Al$_2$O$_3$(0001) | 250°C | 2 | 98 ± 6 | - |
| Al$_2$O$_3$(0001) | 250°C with anneal | 2 | 88 ± 3 | - |

*\* Effective conductance, including thermal transport through film thickness and interfaces.*

Films grown from polyborazylene are much thinner than the single-step, high temperature growths, making direct measurements of their thermal conductivity impossible as the TDTR penetration depth was two orders of magnitude deeper. Instead, for such thin h-BN films thermal effects were interpreted as changes in an overall interface conductance between the top aluminum transducer layer applied for TDTR analysis and the underlying Al$_2$O$_3$ or Si substrates. Such effective thermal interface conductance incorporates thermal transport in the h-BN film as well as its interface resistances with the substrate and aluminum layer. A similar procedure was recently used with graphene layers.[28] From this perspective, there is a clear trend in Table II that the thinner h-BN films grown at 400°C and 250°C provide better effective thermal interface conductance. In all cases it was several times higher than that of the conductance measured for thicker films produced at 1000°C. The observation of higher effective thermal interface conductances for smaller thickness h-BN films can be understood as a decreased resistance of phonon transport through the film thickness. In addition, polyborazylene growth yields better ordered h-BN films near the substrate surface region as well as lower surface roughness values, which would be expected to reduce phonon scattering at the h-BN interfaces. It remains to be determined how crystallinity within the h-BN film and its structural alignment at the interfaces affects thermal conductance. As one indicator, a well ordered 5 nm thick h-BN film produced by direct growth on a copper foil (see supplemental Figure S7) and transferred to sapphire shows an effective thermal interface conductance of 101 ± 5 MWm$^{-2}$K$^{-1}$. This is higher than values in Table II, indicating that ordering within the h-BN film may contribute to the observed improvement of the effective thermal conductance with the polyborazylene growth. Despite the improvement in thermal conductivity observed for the amorphous BN films, the values are still below bulk h-BN; indicating that films synthesized *via* the polyborazylene to BN conversion process presented here are likely more suitable for electronic and optoelectronic applications that do not require thermal management considerations (i.e. low power electronics, sensors).

To understand the impact of h-BN as a substrate on the transport properties of graphene, it is instructive to understand the strain and Fermi velocity reduction which h-BN induces in graphene overlayers. Figure 4 shows the Raman 2D vs. G peak positions for CVD graphene transferred to various dielectric substrates. Vector analysis correlating graphene's 2D ($\omega_{2D}$) and G ($\omega_G$) peak positions to tensile and compressive strain ($e_T$ and $e_C$, respectively), Fermi velocity reduction ($e_{FVR}$), and hole doping ($e_H$), is also shown and is based on the work of Lee *et al.*[29] and Ahn *et al.*[30] Nominally, strain-free and charge-neutral graphene Raman peaks are located at $\omega_G$ and $\omega_{2D}$ peak positions of 1581.6cm$^{-1}$ and 2676.9cm$^{-1}$, respectively, as indicated by the light blue circle in Figure 4.[29] Giovannetti *et al.*[31] have shown that even weak van der Waals forces between graphene and its dielectric environment can lead to perturbation of graphene's electronic band structure through breaking of symmetry between neighboring C atoms in graphene. The modulation of the electronic structure can manifest as a reduction in the Fermi velocity ($v_F$) of charge carriers in graphene.[30] This Fermi velocity reduction (FVR) was found to only effect the 2D peak of graphene while the G peak of graphene, originating from the $E_{2G}$ phonon, is not affected to a first-order approximation.[30,32] Therefore, any reduction in $v_F$ of graphene will manifest as an upward shift in the 2D peak along the $e_{FVR}$ vector shown in Figure 4, where the values shown indicate $\Delta v_F/v_F$ (%) and are negative to indicate a decrease from the intrinsic $v_F$ of graphene ($\approx 10^6$ m/s).[30,32] Raman spectroscopy is also useful in characterizing strain ($\varepsilon$) in a wide variety of materials, since changes in lattice constant due to strain will result in alterations to the respective phonon frequencies.[33] Lee *et al.*[29] then extracted the contribution of strain ($\varepsilon$) for given points in the $\omega_G$-$\omega_{2D}$ space, forming a simple vector model for $e_T$ and $e_C$, as explained elsewhere.[29] Finally, the vector associated with hole doping ($e_H$) was formed from *in-situ* Raman analysis data obtained by Das *et al.*,[34] where hole doping was controlled through electrical gating. Since both electron and hole doping will lead to an increase in $\omega_G$,[29] the results presented in Figure 4 show that the contribution of charge doping to the Raman spectra of the samples investigated here is negligible. Therefore, the $\omega_G$ and $\omega_{2D}$ positions of the transferred CVD graphene investigated here can be vector decomposed exclusively into contributions from strain ($e_T$, $e_C$) and Fermi velocity reduction ($e_{FVR}$) as a function of supporting dielectric substrate.

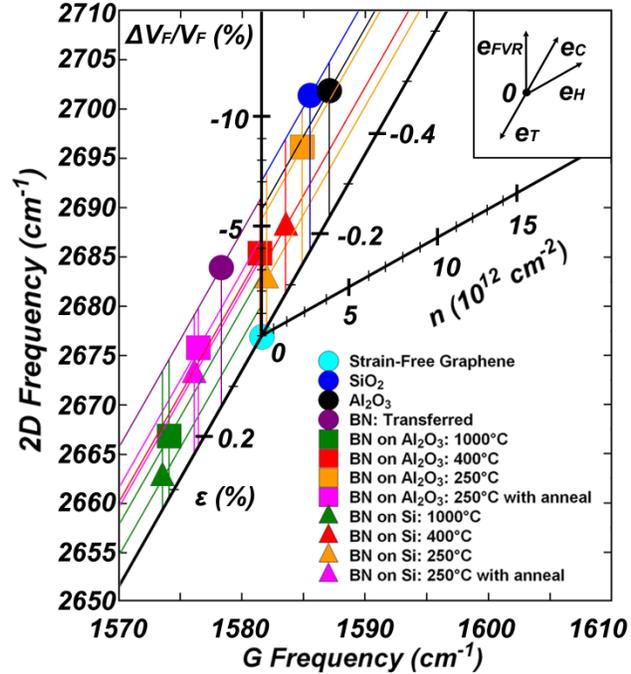

**FIG. 4:** Raman 2D frequency vs. G frequency of CVD graphene transferred to various substrates. The data is vector decomposed to correlate peak shifting to tensile and compressive strain ($e_T$ and $e_C$, respectively), Fermi velocity reduction ($e_{FVR}$), and hole doping ($e_H$), using methods reported by Ahn et al.[30] Inset: expected trajectories of 2D and G peak positions affected by $e_T$, $e_C$, $e_{FVR}$, and $e_H$.

Significant variation in strain and FVR can occur when incorporating graphene with dielectric substrates. Graphene transferred to the bare $SiO_2$ (blue circle) and $Al_2O_3$ (black circle) substrates both experienced compressive stress (-0.17% and -0.24%, respectively) and large FVR (-7.1% and -5.8%, respectively). The compressive stress likely arises from strong adhesion of the graphene to these dielectrics. It has been shown previously that exfoliated graphene on $SiO_2$ substrates experience strong compressive stress and deformation due to ultra-strong adhesion forces with the underlying substrate.[35] It is possible that graphene may have strong adhesion to $Al_2O_3$ as well, based on the shift of $\omega_G$-$\omega_{2D}$ along $e_C$. These strong adhesion forces also perturb the electronic band-structure of the transferred graphene,[35] which again manifests as FVR and results in an upward shift in $\omega_{2D}$. h-BN films grown directly on $Al_2O_3$ (square symbols) and Si (triangular symbols) at 1000°C without a polyborazylene deposition (green symbols) and at 250°C with a pre-growth anneal (pink symbols) resulted in extensive tensile strain in the transferred graphene, with an average of 0.34% and 0.24% tensile strain, respectively. Here, the tensile strain likely results from the transferred graphene layer being stretched across the rough morphology. Interestingly, the FVR for these samples was low compared to the other growth conditions, where the samples grown at 1000°C without a polyborazylene deposition and at 250°C with a pre-growth anneal resulted in a FVR of -2.0% and -4.0%, respectively. This is particularly unexpected since the h-BN samples grown at 1000°C showed the highest degree of crystallinity, as observed with TEM in Figure 4a. Ahn *et al.*[30] showed that exfoliated h-BN supporting dielectrics resulted in an increase in FVR compared to amorphous $SiO_2$ due to the crystalline nature of h-BN, which would induce periodic perturbations to the graphene band structure. However, here the 1000°C h-BN sample shows lower FVR than the amorphous BN samples (grown with the polyborazylene deposition). This again may be attributed to the high surface roughness of these samples. Since the graphene film would be partially suspended over the rough h-BN surface, a large percentage of the graphene film would not be in contact with the h-BN surface, reducing the van der Waals interactions between the graphene and substrate. In this case, the tensile strain resulting from the partial suspension over the substrate would primarily contribute to the shift in Raman peak position. The samples grown at 250°C with a pre-growth anneal show higher FVR despite the fact that they are amorphous and would be expected to contribute less van der Waals interactions. However, this sample was found to be 72% less rough than the 1000°C sample (see Figure 1), and thus would be in greater contact with the h-BN surface, thereby interacting with the transferred graphene to a higher degree. Similarly, the ~5 nm thick h-BN film grown on a Cu foil and transferred to $Al_2O_3$ (purple circle) also induces tensile stress (0.14%) to the graphene layer, due to a RMS surface roughness of 1.5 nm (see supplemental Figure S2c), which is similar to the roughness of the sample grown at 250°C with an anneal. However, the transferred h-BN film shows significant FVR compared to the direct grown h-BN samples. This may be a result of the improved crystallinity, where these films were found to have much higher crystallographic texturing and domain sizes compared to the direct growth h-BN films (see supplemental Figure S6), and thus may not only induce strain from surface roughness but also induce periodic perturbations to the graphene band structure over several nm ranges.

Boron Nitride films grown with a polyborazylene deposition of 250°C *without* a pre-growth anneal (orange symbols) were found to result in the most variation in $\omega_G$ and $\omega_{2D}$ peak positions between the $Al_2O_3$ (square) and Si (triangle) substrates. The sample grown on Si resulted in a negligible compressive strain of -0.02% and a low FVR of -2.2%. The amorphous structure of this film is expected to induce negligible van der Waals interactions to the graphene layer, thus resulting in minimal FVR. However, it should be noted that this film experienced surface pitting (see Figure 1); therefore the Raman line scan for this sample was taken in a region where surface pitting was not observed locally. Although this sample resulted in $\omega_G$ and $\omega_{2D}$ positions comparable with strain-free charge-neutral graphene, this sample is unfortunately unsuitable for graphene device development due to the surface pitting observed across the film surface. In contrast, the BN sample grown on $Al_2O_3$ resulted in a compressive strain of -0.14% and a FVR of -5.4%. It is not fully understood why this sample shows such contrast to the sample grown on Si under the same conditions, as the relatively high FVR is not expected for an amorphous material. However, based on XPS results shown in Table I, it is known that this film is much different chemically than the film grown on Si. Therefore, the increase in FVR may be due to the fact that this sample was found to have significant oxygen contamination. Substitutional oxygen bonding would induce a high density of dangling bonds due to an extra valence electron compared to nitrogen; therefore the increased density of dangling bonds likely induces electrostatic perturbations to graphene's band structure and exacerbates the FVR.

The samples grown at a polyborazylene deposition of 400°C (red symbols) show negligible strain and low FVR for both BN films grown on $Al_2O_3$ (squares) and Si (triangles), where the strain was 0.01% and -0.08% and the FVR was -3.7% and -3.1% for these films, respectively. The amorphous nature of these films results in minimal van der Waals interaction with the graphene layer and minimally degrades the Fermi velocity. Additionally, the presence of both B-B bonding and $BN_xO_y$ bonding (as observed with XPS in Figure 2b) likely results in an overall compensation of valence; resulting in negligible electrostatic interaction from free electrons, further leading to a low FVR. Even more interesting is that the strain induced into graphene is nearly negligible and comparable to strain-free graphene, likely due to low adhesion between graphene and BN as well as the low RMS surface roughness of these films (as shown in Figure 1c and d). These results suggest that the amorphous BN grown with a polyborazylene deposition of 400°C may be an ideal substrate for CVD graphene integration.

Hall transport provides evidence that substrate choice and h-BN properties have significant impact on graphene electronic properties. Figure 5 summarizes carrier mobility as a function of charge carrier density for CVD graphene transferred to various substrates. CVD graphene was transferred to direct growth h-BN on $Al_2O_3$, a 5 nm h-BN film grown on Cu and transferred to $Al_2O_3$, a bare $Al_2O_3$(0001) substrate, and a bare 300 nm $SiO_2$ substrate. Graphene transferred to bare $Al_2O_3$(0001) (orange circles) shows an average mobility of 4218 (±849) $cm^2$/V-s at a hole concentration of $1.9x10^{12}$ (±$4.5x10^{11}$) $cm^{-2}$, while the graphene sample transferred to the $SiO_2$ substrate (brown hexagons) shows an average mobility of 2559 (±124) $cm^2$/V-s at a hole concentration of $3.5x10^{12}$ (±$3.5x10^{11}$) $cm^{-2}$. Both $Al_2O_3$ and $SiO_2$ have low surface optical phonon energies, reported at 55 and 60 meV, respectively.[36] Therefore, both substrates are expected to experience significant phonon-electron coupling, resulting in surface optical phonon scattering. In a separate work, however, it was found that impurity scattering is the dominant scattering mechanism for h-BN gated epitaxial graphene devices, accounting for >90% of the total scattering processes for h-BN coated epitaxial graphene. Though the samples detailed here show reduced carrier densities compared to epitaxial graphene, these samples are likely similarly dominated primarily by impurity scattering.[8] Given that the dielectric constants of $Al_2O_3$ and $SiO_2$ are 12.5 and 3.9,[36] respectively, the higher dielectric constant of $Al_2O_3$ will act to screen charged impurities that would induce Coulombic scattering to graphene, as described by Konar *et al.*[36] Therefore, the effective impurity density of the graphene/$Al_2O_3$ system is reduced compared to graphene/$SiO_2$, leading to higher mobilities for the graphene/$Al_2O_3$ system. Based on temperature dependent Hall Effect measurements and subsequent modeling described by Hollander *et al.*,[8] it was found that the effective impurity concentration is approximately $5.5 \times 10^{11}$ $cm^{-2}$ for a graphene/$SiO_2$ system with the mobilities and sheet densities shown in Figure 5a. Unfortunately, the impurity density of the graphene/$Al_2O_3$ system cannot be extracted without temperature-dependent Hall mobility analysis. The variation in mobility for these two samples can be effectively described by impurity scattering, since surface roughness scattering can be ruled out as both substrates display a similar surface roughness (see supplemental Figure S3). Additionally, the compressive strain and high Fermi velocity reduction (FVR) observed with Raman spectroscopy was similar for these two samples (Figure 4) and likely further limits the mobility for both substrates.

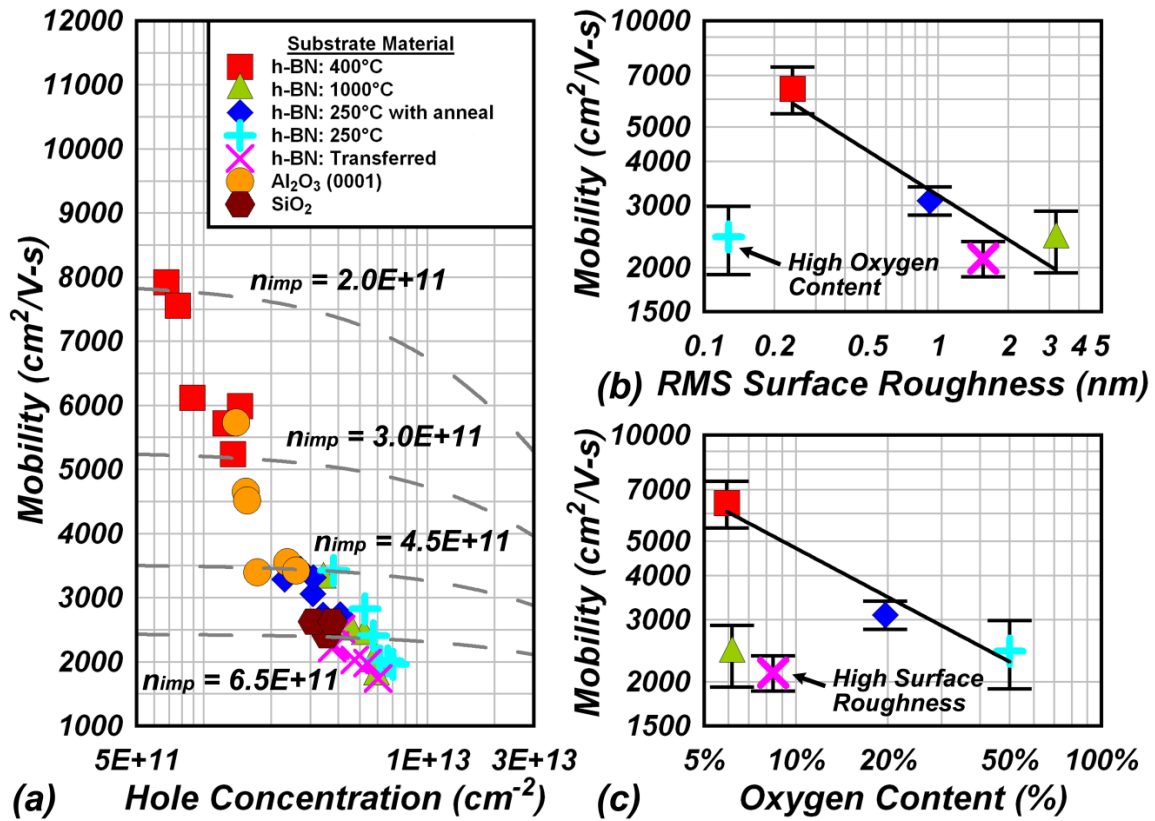

**FIG. 5:** (a) Hall mobility vs. hole concentration of CVD graphene transferred to h-BN dielectrics synthesized at various growth conditions, as well as bare $Al_2O_3$ and $SiO_2$ substrates. Dotted curves indicate the simulated dependency of mobility on carrier density at various impurity densities based on modeling for the h-BN/graphene system described by Hollander *et al.*[8] (b) Log-log plot of graphene Hall mobility *vs*. RMS surface roughness of the underlying h-BN dielectric, as measured by AFM. (c) Log-log plot of graphene Hall mobility *vs*. oxygen content of the underlying h-BN dielectric, as measured by XPS. The symbols shown in (b) and (c) correspond to the legend given in (a).

The dotted lines in Figure 5a represent the simulated dependency of mobility on carrier density for given impurity densities of the h-BN/graphene system.[8] This model assumes an acoustic deformation potential of 4.8 eV[37] and an optical deformation potential of 25.6 eV/A[37] for the CVD graphene and that the surface optical phonon modes (taken as 160 meV) of all h-BN films are equal.[8] These simulated trends indicate the observed variation in mobility between graphene transferred to the various h-BN samples cannot be described simply by changes in carrier density. Instead, based on this modeling, the change in mobility as a function of h-BN growth conditions can be effectively described by impurity concentration. The graphene devices on h-BN films grown at an initial polyborazylene deposition temperature of 400°C (red square) resulted in an average mobility of 6425 (±975) cm$^2$/V-s at a hole concentration of 1.1x10$^{12}$ (±3.0x10$^{11}$) cm$^{-2}$ and reached a maximum mobility of nearly 8000 cm$^2$/V-s. These mobilities correspond to impurity densities ranging from 2.0x10$^{11}$ cm$^{-2}$ to 3.0x10$^{11}$ cm$^{-2}$. In contrast, the h-BN films grown at the various other growth conditions result in reduced mobility and increased hole density, leading to increased impurity densities ranging from 4.5x10$^{11}$ cm$^{-2}$ to 6.5x10$^{11}$ cm$^{-2}$. In other works,[7,8] we found that the benefits of h-BN dielectrics on epitaxial graphene were heavily dependent on the charged impurity density, where impurity concentrations >5x10$^{11}$cm$^{-2}$ resulted in a charged impurity dominated scattering regime where high-k dielectrics outperformed h-BN due to their ability to screen Coulombic scatterers. On the other hand, for impurity concentrations <5x10$^{11}$cm$^{-2}$, surface optical phonon scattering begins to contribute more significantly to the total scattering rate. In this scattering regime, the increased phonon-electron coupling induced by high-k dielectrics (having low surface optical phonon modes) limit the mobility of charge carriers in graphene, while h-BN gated devices benefit from h-BN's high surface optical phonon modes. Therefore the work presented here, as shown in Figure 5a, agree with previous reports, where the high-k dielectric Al$_2$O$_3$ outperforms h-BN samples with high impurity densities, while the h-BN samples with low impurity densities (grown at 400°C) outperform Al$_2$O$_3$.

Unfortunately, the model used for impurity density extraction in Figure 5a does not take into account surface roughness scattering, where significant surface roughness variation between h-BN growth conditions was observed in Figure 1. Increased interfacial surface roughness has not only been shown in literature to directly induce scattering in graphene;[38] but also induces strain in graphene (as validated with Raman spectroscopy in Figure 4), which has been shown to further reduce mobility.[39,40] Therefore, the average mobilities of the transferred graphene on the various h-BN samples is plotted against RMS surface roughness, as measured with AFM, in Figure 5b. A trend is observed for four of the five samples; however the sample grown at 250°C without a pre-growth anneal results in a low average mobility (2445 (±535) cm$^2$/V-s at a hole concentration of 5.8x10$^{12}$ (±1.1x10$^{12}$) cm$^{-2}$) despite the fact that this sample displayed the lowest RMS roughness (130 pm) of all the h-BN samples. To understand this, the average mobilities were also plotted against the oxygen concentration of the h-BN films, as measured with XPS, in Figure 5c. It is likely that substitutional oxygen bonding contributes significantly to the overall impurity density of the graphene/dielectric system since oxygen possesses an extra valence electron than nitrogen, resulting in an increased density of dangling bonds that would induce Coulombic scattering to graphene. Therefore, the oxygen concentration of the h-BN films can be correlated directly to impurity density. From Figure 5c, graphene's mobility is found to decrease with increasing oxygen content, where the sample grown at 250°C without a pre-growth anneal (light blue cross) is dominated by impurity scattering due to its high concentration of substitutional oxygen impurities. Again however, outliers in the observed trend are found. The sample grown at 1000°C without a polyborazylene deposition (green triangle) was found to have an average mobility of 2407 (±478) cm$^2$/V-s at a hole concentration of 4.9x10$^{12}$ (±8.3x10$^{11}$) cm$^{-2}$, while the transferred h-BN sample (pink "X") was found to have an average mobility of 2123 (±243) cm$^2$/V-s at a hole concentration of 4.6x10$^{12}$ (±8.3x10$^{11}$) cm$^{-2}$, despite the fact that these samples had relatively low oxygen impurity concentrations. The results of Figures 5b and 5c suggest that two scattering mechanisms contribute to graphene transport limitations when transferred to the h-BN films. Scattering can be dominated by surface roughness (and strain) as evident in the rough h-BN sample grown at 1000°C (RMS roughness >3 nm) and the transferred h-BN sample (RMS roughness >1.5 nm), or by impurity density as evident by the ultra-smooth but highly oxygen-rich (>50% oxygen content) sample grown at 250°C without a pre-growth anneal. Another option is a combination of the two scattering mechanisms, as evident by the h-BN film grown at 250°C with a pre-growth anneal (blue diamond), where the mobility (3098 (±281) cm$^2$/V-s at a hole concentration of 3.1x10$^{12}$ (±5.6x10$^{11}$) cm$^{-2}$) was found to similarly lie at intermediate points in the trend lines of both Figures 5b and 5c. In this case, the moderate roughness and moderate oxygen contamination lead to contributions from both scattering mechanisms. Further work is required to model the exact impurity concentrations of these samples while taking into account the effects of surface roughness scattering. Based on these results, however, it is apparent that when h-BN is synthesized under conditions that result in both low surface roughness and low oxygen contamination, the beneficial effects of h-BN's high energy surface optical phonon modes can be realized. This was achieved in the present work for h-BN synthesis at an initial polyborazylene deposition temperature of 400°C; where an RMS surface roughness of 240 pm resulted in nearly strain-free transferred graphene (as observed with Raman spectroscopy in Figure 4), while a low oxygen content of <6 atomic % resulted in low impurity densities (<3x10$^{11}$cm$^{-2}$) which allow h-BN's high energy surface optical phonon modes to benefit graphene devices. With use of the 400°C polyborazylene to BN synthesis conditions, transferred CVD graphene devices resulted in an increase in average mobility of 52% and 151% and a reduction in average hole doping of 75% and 224% over the Al$_2$O$_3$ and SiO$_2$ substrates, respectively, with a maximum mobility reaching nearly 8000 cm$^2$/V-s. These results also suggest that even use of nearly amorphous BN, as shown with cross-sectional TEM in Figure 3, can benefit graphene electronics.

**CONCLUSIONS**

Boron nitride was deposited via CVD directly on Si and sapphire substrates using a polyborazylene to BN conversion process. This direct-growth process allows for the use of h-BN as a supporting substrate to CVD-grown graphene. Initial results show that by using a two-step growth process involving the low temperature deposition of a polyborazylene film and the subsequent conversion to BN through a high temperature anneal, ultra-smooth h-BN films down to ~130 pm RMS roughness can be successfully deposited with a uniform thickness controllable down to a few nanometers. This surface roughness is the lowest reported roughness of non-exfoliated BN and represents a >25x reduction in RMS roughness over h-BN films grown without the polyborazylene to BN conversion. Control of the stoichiometry appears to be difficult during the polyborazylene to BN conversion, where boron-rich films were obtained at polyborazylene deposition temperatures of 250°C and 400°C. Additionally, these films were found with TEM to be nearly amorphous with only short-range order. Films produced with initial polyborazylene growth at reduced temperatures show improved thermal interface conductance as compared to the films directly grown at 1000°C, which was attributed to their decreased thickness, lower surface roughness, and better interface structural alignment thereby decreasing phonon scattering. The overall low interfacial thermal conductances for these films indicates that these dielectrics are likely more suitable for low-power graphene applications, however. Raman spectroscopy was utilized to quantify changes in strain and Fermi velocity reduction (FVR) of transferred CVD graphene as a function of the dielectric underlying substrate. It was found that the films with a high surface roughness resulted in significant tensile strains and that the films on bare $SiO_2$ and $Al_2O_3$ resulted in strong compressive strains and high FVR (likely from high adhesion strengths). The amorphous BN films grown with an initial polyborazylene deposition of 400°C resulted in nearly strain-free graphene, due to its low surface roughness and low adhesion, as well as low FVR as a result of negligible van der

Waals interactions (due to its amorphous structure and valence compensated chemistry) with the overlying graphene. These results suggest that these films may be excellent candidates for CVD graphene integration as a supporting dielectric. Indeed, when used as a supporting substrate, the films grown at a polyborazylene deposition temperature of 400°C showed the best potential for improved transport properties of transferred CVD graphene, resulting in a 52% and 151% increase in mobility compared to CVD graphene on bare sapphire and $SiO_2$ substrates, respectively. The extracted impurity density of $<3\times10^{11} cm^{-2}$ for graphene transferred to this sample suggests surface optical phonon scattering contributes significantly to the overall scattering rate, where h-BN excels compared to high-k dielectrics due to its high surface optical phonon modes. Importantly, these results suggest that even amorphous/turbostratic BN can benefit graphene electronics.

**METHODS**

Hexagonal boron nitride is grown on single crystal $Al_2O_3$(0001) and Si(111) in a 80mm diameter horizontal tube furnace *via* a thermal CVD method utilizing a single ammonia borane ($NH_3BH_3$) precursor (Sigma-Aldrich, part #682098). The $Al_2O_3$ and Si substrates are cleaned with acetone, isopropyl alcohol, and Nanostrip 2x prior to growth to remove organic and ionic contaminants. The Si substrates are then placed in a 10:1 BOE HF solution to remove the native oxide. Solid ammonia borane powder is sublimed at 135°C and transported into the tube furnace by an $H_2$/Ar carrier gas (5% of total flow rate). Growth occurs at 100 Torr with growth times ranging between 5-120 minutes, depending on desired film thickness and polyborazylene deposition temperature. After the growth procedure, the ammonia borane carrier gas is turned off and the furnace is allowed to slowly cool to room temperature in a

250mTorr Ar/H$_2$ environment. Graphene was prepared via a catalytic CVD method on 25µm 99.999% purity Cu foils at 1050°C, 1 Torr

The as-grown h-BN films are characterized using atomic force microscopy (AFM), scanning electron microscopy (SEM), x-ray photoelectron spectroscopy (XPS), optical ellipsometry, and cross-sectional transmission electron microscopy (TEM). A Bruker Icon AFM with a scan rate of 0.5 Hz and a resolution of 512 points per line was utilized for AFM measurements. A Leo 1530 field emission SEM with an accelerating voltage of 5kV was used to acquire SEM micrographs. A Kratos Axis Ultra XPS system utilizing an Al kα source with energy of 1486 eV was used for XPS analysis. A Gaertner L116C variable angle ellipsometer was used for thickness measurement of the as-grown h-BN films assuming a bulk refractive index of 1.67. A JEOL 2010F TEM was used for cross-sectional TEM analysis of h-BN films on sapphire and Si(111). Cross-plane thermal transport properties of thin h-BN films were measured using time-domain thermoreflectance (TDTR), which is a femtosecond laser pump-probe based approach to monitor time-resolved, temperature-induced changes in optical reflectivity.[25] For TDTR preparation, each sample was coated with ~70 nm of aluminum metal to serve as a thermo-reflectance transducer. Analysis of TDTR data was performed using a nonlinear least-squares application to Cahill's frequency domain model.[41] For this analysis, the physical properties of the aluminum layer were held constant. The aluminum thickness was quantified by picosecond acoustics and bulk aluminum thermal properties were assumed. For data fitting to the thermal diffusivity model, heat capacities of bulk sapphire, silicon, and h-BN were used. For each sample, data sets were statistically analyzed to obtain average ± standard deviation values. Raman spectroscopy was used to investigate strain and van der Waals interactions that lead to Fermi velocity reduction of transferred graphene on various dielectrics. Raman data was obtained through 20µm line scans utilizing a WiTec CRM200 at an incident laser wavelength of 488 nm. Hall mobility and hole concentration of transferred CVD graphene was obtained with a Nanometrics 4-point probe Hall mobility measurement system. van der Pauw test structures with

10x10μm$^2$ Hall crosses and Ti/Au (10/50 nm) contacts were prepared via standard UV photolithography techniques.

**ACKNOWLEDGMENTS**

This work was supported by the Office of Naval Research, Contract N00014-12-C-0124 , and the Defense Threat Reduction Agency, Contract HDTRA1-10-1-0093. Any opinions, findings, conclusions, or recommendations expressed in this material are those of the authors and do not necessarily reflect the views of the sponsors. Support for the WiteC Raman system, Leo 1530 SEM, Bruker Icon AFM, Kratos Axis Ultra XPS, JEOL 2010F TEM, and nanofabrication facility was provided by the National Nanotechnology Infrastructure Network at Penn State. Funding for M. Bresnehan provided by the Applied Research Laboratory (ARL) Exploratory and Foundational research assistantship.